\documentclass[prd,eqsecnum,showpacs,aps]{revtex4}
\usepackage{latexsym}
\usepackage{graphicx}
\begin{document}
\title{Second-order gravitational effects of local inhomogeneities \\
on CMB anisotropies and non-Gaussian signatures}

\author{Kenji Tomita}
\affiliation{Yukawa Institute for Theoretical Physics, 
Kyoto University, Kyoto 606-8502, Japan}
\date{\today}

\begin{abstract}
Based on the second-order nonlinear theory of perturbations in
non-zero $\Lambda$ flat cosmological models, we study the gravitational
effects of local inhomogeneities on cosmic microwave
background (CMB) anisotropies. As the local inhomogeneities we
consider firstly large-scale dipole and quadrupole distributions of
galaxies around us and next an isolated cluster-scale matter
distribution. It is found that, due to the second-order integral
Sachs-Wolfe effect, the north-south asymmetry of CMB
anisotropies and non-Gaussian signatures (in terms of scale-dependent
estimators of kurtosis) in a spot-like object are caused from these 
matter distributions along light  
paths. Our theoretical results seem to be consistent with recent
various observational results which have been shown by Hansen et al.,
Eriksen et al., Vielva et al. and Cruz et al.    
\end{abstract}
\pacs{98.80.-k, 98.70.Vc, 04.25.Nx}

\maketitle


\section{Introduction}
\label{sec:level1}

In the modern precise cosmology, the observations of cosmic microwave
background (CMB) anisotropies are bringing us important informations
on the structure of the universe \cite{map,spg,komt}. 
To analyze them, the theories not only of first-order perturbations but
also of second-order perturbations are necessary and useful \cite{rev}.
In a previous paper\cite{tom} we studied the general behavior of
relativistic 
second-order perturbations in non-zero $\Lambda$ flat cosmological
models, which correspond to the first-order scalar perturbations, and
derived the basic equation for the second-order integral Sachs-Wolfe
effect of CMB anisotropies. In the subsequent second previous paper
\cite{tom1} we treated a case when the
first-order perturbations consist of primordial random density
perturbations and a local inhomogeneity which does not belong to the
former perturbations, and derived the second-order temperature
perturbations caused by the coupling of the above two types of
perturbations. It was found as a result that the nonlinear behavior of
the latter temperature perturbations may explain the north-south
asymmetry of CMB anisotropies observed by Eriksen et
al.\cite{erk1,erk2} and Hansen et al.\cite{hans1,hans2,hans3}, 
 when we assume a large-scale
matter distribution with dipole component around us as a local
inhomogeneity.  

In this paper, we show in Section II the derivation of power spectra, 
the possible expansions of anisotropies with spherical harmonics, and
the scale-dependent estimator
representing non-Gaussian signatures, and consider in Section III a
large-scale matter distribution with dipole and quadrupole components
around us to bring more realistic north and south asymmetry of CMB
anisotropies. To consider non-Gaussian signatures in the spot-like
object (which was observed by Vielva et al.\cite{viel} and Cruz et
al.\cite{cruz}), moreover, we take up in Section IV  
a cluster-scale matter distribution which is so isolated to 
be not included in the primordial random density perturbations, and
study its second-order gravitational influence on CMB anisotropies. 
No non-Gaussian signal is found in the original definition of
skewness and kurtosis, but it is shown that a non-Gaussian signal
similar to the observed one can be derived in the form of a
scale-dependent estimator of kurtosis. 
 Concluding remarks follow in Section V.

\section{CMB anisotropies with second-order nonlinearity}
\label{sec:level2}

\subsection{Power spectra of CMB anisotropies}

The spatially flat background model is assumed and its metric is
 expressed as
\begin{equation}
  \label{eq:ma1}
 ds^2 =  g_{\mu\nu} dx^\mu dx^\nu = 
a^2 (\eta) [- d \eta^2 + \delta_{ij} dx^i dx^j],
\end{equation}
where the conformal time $\eta (=x^0)$ is related to the cosmic time
$t$ by $dt = a(\eta) d\eta$, the Greek and Latin letters denote
$0,1,2,3$ and $1,2,3$, respectively, and
$\delta_{ij} (= \delta^i_j = \delta^{ij})$ are the Kronecker
delta. The matter density $\rho$ and the scale factor $a$ have the
relations 
\begin{equation}
  \label{eq:ma2}
\rho a^2 = 3(a'/a)^2 - \Lambda a^2, \quad {\rm and } \quad 
\rho a^3 = \rho_0,
\end{equation}
where a prime denotes $\partial/\partial \eta$,  $\Lambda$ is the
cosmological constant,  $\rho_0$ is an integration constant and the
units $8\pi G = c =1$ are used.

The first-order and second-order metric perturbations
$\mathop{\delta}_1 g_{\mu\nu} (\equiv 
h_{\mu\nu})$ and $\mathop{\delta}_2 g_{\mu\nu} (\equiv
\ell_{\mu\nu})$, respectively, were derived explicitly by imposing the
synchronous coordinate condition (in \cite{tom})  :
\begin{equation}
  \label{eq:ma3}  
h_{00} = h_{0i} = 0 \quad {\rm and} \quad \ell_{00} = \ell_{0i} = 0.
\end{equation}
Here only first-order perturbations in the growing mode are shown for
the following use. The metric perturbation is
\begin{equation}
  \label{eq:ma4}
h^j_i =  P(\eta) F_{,ij},  
\end{equation}
where $F$ is an arbitrary potential function of spatial coordinates
$x^1, x^2$ and $x^3, \ h^j_i = \delta^{jl}h_{li}$, and $P(\eta)$
 satisfies  
\begin{equation}
  \label{eq:ma5}
P'' + {2a' \over a} P' -1 = 0, 
\end{equation}
The velocity perturbation $\mathop{\delta}_1 u^\mu$ vanishes, and the
density perturbation is expressed as
\begin{equation}
  \label{eq:ma6}
\mathop{\delta}_1 \rho/\rho = {1 \over \rho a^2} \Bigl({a'\over a}P'
-1\Bigr) \Delta F.
\end{equation}

Now let us assume a form of the potential function
\begin{equation}
  \label{eq:m1}
F({\bf x}) = F_P({\bf x}) + F_L({\bf x})
\end{equation}
with
\begin{equation}
  \label{eq:m2}
F_P = \int d{\bf k} \alpha ({\bf k}) e^{i{\bf kx}} \quad {\rm and} \quad
F_L = R(r) g(\theta),
\end{equation}
where $F_P$ is the part of primordial density perturbations with
random variables $\alpha ({\bf k})$ and $F_L$ is the part of local
homogeneities with radial and angular dependences specified by $R(r)$ and
$g(\theta)$. The latter function is expressed using Legendre
polynomials as 
\begin{equation}
  \label{eq:m3}
g(\theta) = \sum_l b_l P_l (\theta)
\end{equation}
and by the averaging process for $\alpha ({\bf k})$, we have   
\begin{equation}
  \label{eq:m4}
\langle \alpha ({\bf k}) \alpha ({\bf k'}) \rangle = (2\pi)^{-2}
{\cal P}_F ({\bf k}) \delta ({\bf k} + {\bf k'}).
\end{equation}

The first-order density perturbation corresponding to local
inhomogeneities is  
\begin{eqnarray}
  \label{eq:m5} 
(\mathop{\delta}_1 \rho/\rho)_L = {1 \over \rho a^2} \Bigl({a'\over
a}P'-1\Bigr) \tilde{R} (r) g (\theta),   
\end{eqnarray}
where
\begin{equation}
  \label{eq:m6}
\tilde{R} (r) = {1 \over r^2}{d \over dr}(r^2 R_{,r}) - \sum_l
l(l+1)b_l r^2 R
\end{equation}
with $R_{,r} \equiv d R/dr$ and $P' \equiv dP/d\eta$. 
The solution for Eq.(\ref{eq:ma5}) is expressed as
\begin{eqnarray}
  \label{eq:m8}
P &=& -{2 \over 3\Omega_0} y^{-3/2} (\Omega_0 +\lambda_0 y^3) \int^y_0
dy' {y'}^{3/2}/\sqrt{\Omega_0 +\lambda_0 {y'}^3} + {2 \over
3\Omega_0}y,\cr
\eta &=& \int^y_0
dy' {y'}^{-1/2}/\sqrt{\Omega_0 +\lambda_0 {y'}^3}, 
\end{eqnarray}
where $y \equiv a/a_0$ is $1$ at the present epoch. In sections III
and IV, we treat the large-angle case  $g(\theta) = \cos \theta +
\beta \cos^2 \theta$ and the small-angle case  $g(\theta) = P_l
(\theta)$ with $l = 20$, respectively. 

In the previous paper \cite{tom1} we derived the temperature
perturbations in the first-order and second-order in the case
$g(\theta) = \cos \theta$. As for the temperature perturbations with
general $g(\theta)$, we have in the first-order and
second-order     
\begin{eqnarray}
  \label{eq:m9}
\mathop{\delta}_1 T/T &=& \Theta_P + \Theta_L g(\theta),\cr
\mathop{\delta}_2 T/T &=& \Theta_{LP} + \Theta_{LL} + \Theta_{PP},
\end{eqnarray}
where $\Theta_{P}$ and $\Theta_{L}$ come from the contributions of
only $F_P$ and $F_L$, respectively, and $\Theta_{PP}, \Theta_{LL}$ and
$\Theta_{LP}$ come from the contributions of only $F_P$, only $F_L$
and the coupling of $F_P$ and $F_L$, respectively. In the following,
$\Theta_{PP}$ is neglected because it is small enough, compared with
$\Theta_{P}$, and $\Theta_{L}$ and $\Theta_{LL}$ are expressed as
\begin{eqnarray}
  \label{eq:m10}
\Theta_L &=& {1 \over 2} \int^{\lambda_e}_{\lambda_o} d\lambda P'
(\eta) R_{,rr}, \cr
\Theta_{LL} &=& A [g(\theta)]^2 +B [g'(\theta)]^2 +C g(\theta)
[g''(\theta)+\cot \theta g'(\theta)] + D [(g''(\theta))^2 +\cot^2
\theta (g'(\theta))^2], 
\end{eqnarray}
where $\lambda$ is the affine parameter, $(\eta, r) = (\lambda,
\lambda_0 - \lambda)$ along the light path, and $\lambda_0$ is an
observer's value at present. The expressions of $A,
B, C$ and $D$ are shown in Appendix. Moreover, $\Theta_{P}$ and
$\Theta_{LP}$ are 
\begin{eqnarray}
  \label{eq:m10a}
\Theta_{P} &=& \int d{\bf k} \alpha ({\bf k}) \sum_l Q_l P_l
(\mu),\cr 
\Theta_{LP} &=& \int d{\bf k} \alpha ({\bf k}) \sum_l \Delta Q_l P_l
(\mu) g(\theta),
\end{eqnarray}
with $\mu = {\bf n}\cdot {\bf k}/k$, where ${\bf n}$ is a directional
unit vector with angles $\theta$ and $\phi$. The coefficients are
defined as 
\begin{eqnarray}
  \label{eq:m10b}
Q_l &=& {1 \over 2} (-1)^l (2l +1) {\cal H}_{P}^{(l)}, \cr 
\Delta Q_l &=& {1 \over 4} (-1)^l (2l +1) {\cal H}_{LP}^{(l)},  
\end{eqnarray}
where the expressions of ${\cal
H}_P^{(l)}$ and ${\cal H}_{PL}^{(l)}$ are the same as Eqs.(3.6) and
(3.12) in the previous paper\cite{tom1}, except for $\Phi$. For the above
$g(\theta)$, we have
\begin{eqnarray}
  \label{eq:m10c}  
\Phi \equiv &-&PP''R_{,rrr}+ \Bigl[2P'''P + {1 \over 2}P' -(P''')_e
P\Bigr] R_{,rr} - {1 \over 2}P''R_{,r} - P''' R \cr
&+& P'''' \int^{\lambda}_{\lambda_o} d\bar{\lambda} PR_{,rr} 
-{1 \over 7}(3P'P''+PP''')[4R_{,rr}+9R_{,r}/r -(6\gamma-3) R/r^2] \cr 
&+&{2 \over 7}[(P')^2+PP''][4R_{,rrr}+9R_{,rr}/r -6(2\gamma-1)
R_{,r}/r^2 
+6(2\gamma-1) R/r^3] \cr
&-&{1 \over 7}PP'[4R_{,rrrr}+9R_{,rrr}/r -(12\gamma +3)R_{,rr}/r^2
+18(2\gamma-1) R_{,r}/r^3 -18(2\gamma-1) R/r^4] \cr
&+& {3 \over 7}PP'k^2 [R_{,rr} -R_{,r}/r + (\gamma -1) R/r^2],
\end{eqnarray}
where $\gamma = 2(1+\beta)$ or $l(l+1)$ for $g(\theta) = \cos \theta +
\beta \cos^2 \theta$ or $P_l (\theta)$, respectively.

The power spectra of CMB anisotropies are
\begin{eqnarray}
  \label{eq:m11}
\langle (\mathop{\delta}_1 T/T)^2 \rangle &=& \langle (\Theta_P)^2
\rangle  + (\Theta_L)^2  g^2 (\theta),\cr
\langle \mathop{\delta}_1 T/T \mathop{\delta}_2 T/T \rangle &=& \langle
\Theta_P \Theta_{LP} \rangle + \Theta_L \Theta_{LL} g(\theta) 
\end{eqnarray}
with
\begin{eqnarray}
  \label{eq:m12}
(T_0)^2 \langle (\Theta_P)^2 \rangle &=&  \sum_l {2l+1 \over 4\pi} C_l,\cr
(T_0)^2 \langle \Theta_P \Theta_{LP} \rangle &=&  \sum_l {2l+1
\over 4\pi} \Delta C_l g(\theta),
\end{eqnarray}
where 
\begin{eqnarray}
  \label{eq:m13}
C_l &=&  (T_0)^2 \int dk k^2 {\cal P}_F (k)
|{\cal H}_P^{(l)} (k) |^2, \cr
 \Delta C_l &=& {1 \over 2} (T_0)^2 \int dk k^2 {\cal P}_F (k) {\cal
H}_P^{(l)} {\cal H}_{PL}^{(l)},  
\end{eqnarray}
and $T_0$ is the present CMB temperature.  It is important that
$\langle \Theta_P \Theta_{LP} \rangle$ is proportional to $g(\theta)$
which is the angular component of $F_L$.

\subsection{Expansions of $\Theta_P$ and $\Theta_{LP}$ in terms of
spherical harmonics}

Let us expand $\Theta_P$ and $\Theta_{LP}$ as functions of the
directional unit vector \ ${\bf n} (\theta,\phi)$ as follows:
\begin{eqnarray}
  \label{eq:n1}  
\Theta_P ({\bf n}) &=& \sum_{lm} a_{P,lm}\ Y_{l,m} (\theta,\phi), \cr
\Theta_{LP}  ({\bf n}) &=& \sum_{lm} a_{LP,lm}\ Y_{l,m} (\theta,\phi),
\end{eqnarray}
and consider another expansion 
\begin{equation}
  \label{eq:n2}  
\Theta_{LP} ({\bf n}) = \sum_{lm} b_{LP,lm} \ g(\theta) 
Y_{l,m} (\theta,\phi)
\end{equation}
for comparison. Then we have
\begin{eqnarray}
  \label{eq:n3}
a_{P,lm} &=& \int Y^*_{l,m} (\theta,\phi) \Theta_P ({\bf n}) d
\Omega, \cr
a_{LP,lm} &=& \int Y^*_{l,m} (\theta,\phi) \Theta_{LP} ({\bf n}) d
\Omega, \cr
b_{LP,lm} &=& \int Y^*_{l,m} (\theta,\phi) [\Theta_{LP}({\bf n})
/g(\theta)]  d\Omega,  
\end{eqnarray}
where $d \Omega = \sin \theta d\theta d\phi$. Using the relation 
\begin{equation}
  \label{eq:n4}  
\int^\pi_{-\pi} \int^\pi_0 Y_{l,m} (\theta,\phi) P_l(\mu) \sin \theta
d\theta d\phi = {4\pi \over 2l+1} Y_{l,m} (\theta_k,\phi_k),
\end{equation}
we obtain from Eqs.(\ref{eq:m10a}) 
\begin{eqnarray}
  \label{eq:n5}  
a_{P,lm} &=&  {4\pi \over 2l+1} \int d{\bf k} \alpha ({\bf k}) Q_l
Y_{l,m}^* (\theta_k,\phi_k), \cr 
b_{LP,lm} &=& {4\pi \over 2l+1} \int d{\bf k} \alpha ({\bf k}) \Delta Q_l
Y_{l,m}^* (\theta_k,\phi_k).
\end{eqnarray}
For $a_{LP,lm}$, we consider the case $g(\theta) = \cos \theta + \beta
\cos^2 \theta$  with a constant $\beta$ for example, and use the
relation
\begin{equation}
  \label{eq:n6}  
\cos \theta P^m_l = {1 \over 2l+1}[(l+m)P^m_{l-1} + (l-m+1)P^m_{l+1}].
\end{equation}
Then we obtain
\begin{equation}
  \label{eq:n7}  
a_{LP,lm} = a_{LP,lm}^{(1)} + \beta a_{LP,lm}^{(2)},
\end{equation}
where
\begin{eqnarray}
  \label{eq:n8}  
a_{LP,lm}^{(1)} &=&  {4\pi \over 2l+1} N_{l,m} \int d{\bf k} \alpha
({\bf k}) \Bigl[{l+m \over 2l-1}
{\Delta Q_{l-1} \over N_{l-1,m}} Y^*_{l-1,m} (\theta_k,\phi_k) \cr
 &+& {l-m+1 \over 2l+3}
{\Delta Q_{l+1} \over N_{l+1,m}} Y^*_{l+1,m} (\theta_k,\phi_k)\Bigr] \cr
a_{LP,lm}^{(2)} &=& {4\pi \over 2l+1} N_{l,m} \int d{\bf k} \alpha
({\bf k}) \Bigl\{{(l+m)(l+m-1) \over 2l-1}
{\Delta Q_{l-2} \over N_{l-2,m}} Y^*_{l-2,m} (\theta_k,\phi_k) \cr
 &+& {(l-m+1)(l-m+2) \over 2l+3}
{\Delta Q_{l+2} \over N_{l+2,m}} Y^*_{l+2,m} (\theta_k,\phi_k) \cr
&+& \Bigl[{(l+m)(l-m) \over 2l-1} + {(l-m+1)(l+m+1) \over 2l+3} \Bigr]
{\Delta Q_{l} \over N_{l,m}} Y^*_{l,m} (\theta_k,\phi_k)\Bigr\}
\end{eqnarray}
with $Y_{l,m}(\theta,\phi) = N_{l,m} P^m_l (\theta) e^{im\phi}$ and
\begin{equation}
  \label{eq:n9}  
N_{l,m} \equiv \Bigl[{2l+1 \over 2\pi} {(l-m)! \over (l+m)!} \Bigr]^{1/2}.
\end{equation}

When we compare the spatial average of the product $a_{P,lm}^* \cdot
a_{LP,l'm'}$ and that of the product $a_{P,lm}^* \cdot
b_{LP,l'm'}$, we find a simple expression
\begin{equation}
  \label{eq:n10}  
<a_{P,lm}^* b_{LP,l'm'}> \propto \delta_{l,l'}  \delta_{m,m'} 
\end{equation}
and rather complicated expressions
\begin{eqnarray}
  \label{eq:n11}  
<a_{P,lm}^* a_{LP,l'm'}^{(1)}> &\propto& {\rm terms \ with} \
\delta_{l,l'\pm 1}  \delta_{m,m'} , \cr
<a_{P,lm}^* a_{LP,l'm'}^{(2)}> &\propto& ({\rm terms \ with} \
\delta_{l,l'} 
+ {\rm terms \ with} \ \delta_{l,l'\pm 2} )\ \delta_{m,m'}.
\end{eqnarray}
Moreover, $<a_{P,lm}^* b_{LP,l'm'}>$ is consistent with $\Delta C_l$
in Eq.(\ref{eq:m13}).
The characteristics of north-south asymmetry is, therefore, described
more clearly using $b_{LP,lm}$ than using  $a_{LP,lm}$, though we can
expand the second-order CMB anisotropies in the two forms of
expansions ( Eqs.(\ref{eq:n1}) and (\ref{eq:n2})). 

\subsection{Non-Gaussian signature}

Now let us derive various quantities
representing non-Gaussianity. First we consider the dispersion
$\sigma$. Since $\langle \delta T/T \rangle = \Theta_L g(\theta) +
\Theta_{LL}$, $\sigma$ is defined by
\begin{eqnarray}
  \label{eq:m14}  
\sigma^2 &\equiv& \langle (\delta T/T - \langle \delta T/T \rangle)^2
\rangle     \cr
&=&  \langle (\Theta_P)^2 \rangle^2 + 2 \langle \Theta_P \Theta_{LP}
\rangle g(\theta).   
\end{eqnarray}
Then we can consider the skewness ($S$) and the kurtosis ($K$) of CMB
anisotropies, taking into account the coupling of primordial
perturbations ($P$) and local inhomogeneities ($L$) in the
second-order. According to the ordinary definitions, they are defined by
\begin{eqnarray}
  \label{eq:m15}  
S &\equiv& \langle (\delta T/T - \langle \delta T/T \rangle)^3
\rangle/\sigma^3,     \cr
K &\equiv& \langle (\delta T/T - \langle \delta T/T \rangle)^4
\rangle/\sigma^4 - 3.
\end{eqnarray}
For $(\delta T/T - \langle \delta T/T \rangle)^2 = (\Theta_P)^2 
+ 2\Theta_P
\Theta_{LP} g(\theta)$, it is found that both of them vanish.

Here let us consider moreover the expectation values of various
scale-dependent estimators $\bar{\sigma} (\theta), \ \bar{S} (\theta)$
and $\bar{K} (\theta)$ in the range of $\theta$, which include 
the estimators due to the spherical Mexican hat wavelet (SMHW).
We adopt the northern pole or the center of a cluster-scale as the
fixed point. The expectation values are defined in  
terms of $\mu \equiv \cos \theta$ as  
\begin{eqnarray}
  \label{eq:m16}    
[\bar{\sigma}]^2 &=& {N(\theta_0)}^{-1} \int^1_0 d\mu \sigma^2 \psi
(\theta,\theta_0) \cr
&=& \langle (\Theta_P)^2 \rangle + 2 \langle \Theta_P \Theta_{LP}
\rangle {N(\theta_0)} \int^1_0 d\mu g(\theta)\psi (\theta,\theta_0), 
\end{eqnarray}
where $N(\theta_0)$ is the normalization factor given by $N(\theta_0)
= \int^1_0 \ d\mu \ \psi (\theta,\theta_0)$. For SMHW, we have 
 
\begin{equation}
  \label{eq:m16a}    
\psi (\theta,\theta_0) \propto (1+(xR)^2)^2 (2-x^2)^2 e^{-x^2/2}
\end{equation}
with $x \equiv (2/R) \tan (\theta/2)$\ \cite{mart}, where $R$ is the
size of Mexican hat related to the angular scale $\theta_0$.
The expectation values of skewness and kurtosis estimators are
expressed as  
\begin{equation}
  \label{eq:m17}    
\bar{S} (\theta) = {N(\theta_0)}^{-1} \int^1_0 d\mu \psi
(\theta,\theta_0) \langle (\delta
T/T - \langle \delta T/T \rangle)^3 \rangle/[\bar{\sigma}]^3, 
\end{equation}
\begin{eqnarray}
  \label{eq:m18}   
\bar{K} (\theta) &=& {N(\theta_0)}^{-1} \int^1_0 d\mu \psi
(\theta,\theta_0) \langle (\delta
T/T - \langle \delta T/T \rangle)^4 \rangle/[\bar{\sigma}]^4 -3 \cr
&=& 3 \Bigl\{{N(\theta_0)}^{-1} \int^1_0 d\mu \psi (\theta,\theta_0)
[\langle (\Theta_P)^2 \rangle
 + 2 \langle \Theta_P \Theta_{LP} \rangle g(\theta)]^2
-[\bar{\sigma}]^4 \Bigr\}/ [\bar{\sigma}]^4. 
\end{eqnarray}
While $\bar{S} (\theta)$ vanishes, $\bar{K} (\theta)$ reduces to
\begin{equation}
  \label{eq:m19}    
\bar{K} (\theta) = 12 {\langle \Theta_P \Theta_{LP} \rangle^2 \over
[\bar{\sigma}]^4} \Bigl\{{N(\theta_0)}^{-1} \int^1_0 d\mu \psi
(\theta,\theta_0) [g(\theta)]^2 - \Bigl({N(\theta_0)}^{-1} \int^1_0 d\mu
\psi (\theta,\theta_0) [g(\theta)] \Bigr)^2 \Bigr\} 
\end{equation}
with $\mu = \cos \theta$. Here it is to be noticed that, when the
third-order terms are taken into account, we have $\mathop{\delta}_3
T/T = \Theta_{LPP} + \Theta_{LLL} +\Theta_{PLL} +\Theta_{PPP}$ and the
terms such as $\langle (\Theta_P)^2 \rangle\langle \Theta_P
\Theta_{PLL}\rangle$ seem to have contributions (to $\bar{K}$)
comparable with $(\langle \Theta_P \Theta_{LP} \rangle )^2$, but it is
found that they are canceled and disappear.

\section{Large-scale matter distributions with dipole and quadrupole
symmetries} 
\label{sec:level3}

In order to study the gravitational effect of a more realistic
large-scale structure around us than that in our previous paper, we
consider a model with dipole and quadrupole symmetries, in which the
angular part of $F_L ({\bf x}) \ (= R(r) g(\theta))$ is
\begin{equation}
  \label{eq:b1}    
g(\theta) = \cos \theta + \beta \cos^2 \theta = {1 \over 3}[P_0(\mu) +
3P_1(\mu) +2P_2(\mu)]
\end{equation}
with $\mu = \cos \theta$ and $P_l(\mu)$ is the Legendre
polynomial. For $R(r)$ we use the following four types of functions as in
the previous paper\cite{tom1}: 
\begin{eqnarray}
  \label{eq:b2}
R =&& R_0 \exp[-\alpha (x-1)^2], \quad {1 \over 2} R_0 [1 + \cos 2\pi
(x-1)], \cr
&& R_0 x^2 \exp[-\alpha (x-1)^2], \ {\rm and} \ {1 \over 2} R_0 x^2 [1
+ \cos 2\pi (x-1)]
\end{eqnarray}
in the interval $x = [x_1, x_2]$ with $x \equiv r/r_c$, in which
$(x_1, x_2) \equiv (r_1,r_2)/r_c = (0.5,1.5)$ and $a_0 r_c$ is
$\approx 300 h^{-1}$Mpc ($H_0 = 100h$ km/s/Mpc).  In all types we
have $R = 0$ for $x > x_2$ or $x < x_1$. $R_0$ is the normalization
constant and a constant $\alpha$ is chosen as 20. 

The above functions were chosen so that they have radially convex 
behaviors and the first two of
them are symmetric for $x > 1$ and $x < 1$, while the last two have
small asymmetry. 

The powers of CMB
anisotropies are calculated using Eqs.(\ref{eq:m12}) and (\ref{eq:m13})
for $l = 1 - 22$. The values of $l(l+1) C_l \xi$ and $l(l+1) \Delta
C_l \xi/R_0$ for $\beta = -0.3$ are shown in Table I, in which G,S, MG
and MS represent the Gaussian type (G), the sine type (S), the
modified Gaussian type (MG) and the modified sine type (MS). Here
$\xi \equiv 2\pi/[{\cal P}_{F0} (T_0)^2]$ and ${\cal P}_{F0} =2.1
\times 10^{-8}$.

\begin{figure}[h]
\caption{\label{fig:erik} The $l$ dependence of $l(l+1) [C_l + 2\delta
C_l g(\theta)]$ in the northern pole ($\theta=0$) and southern pole
($\theta=\pi$).} 
\includegraphics{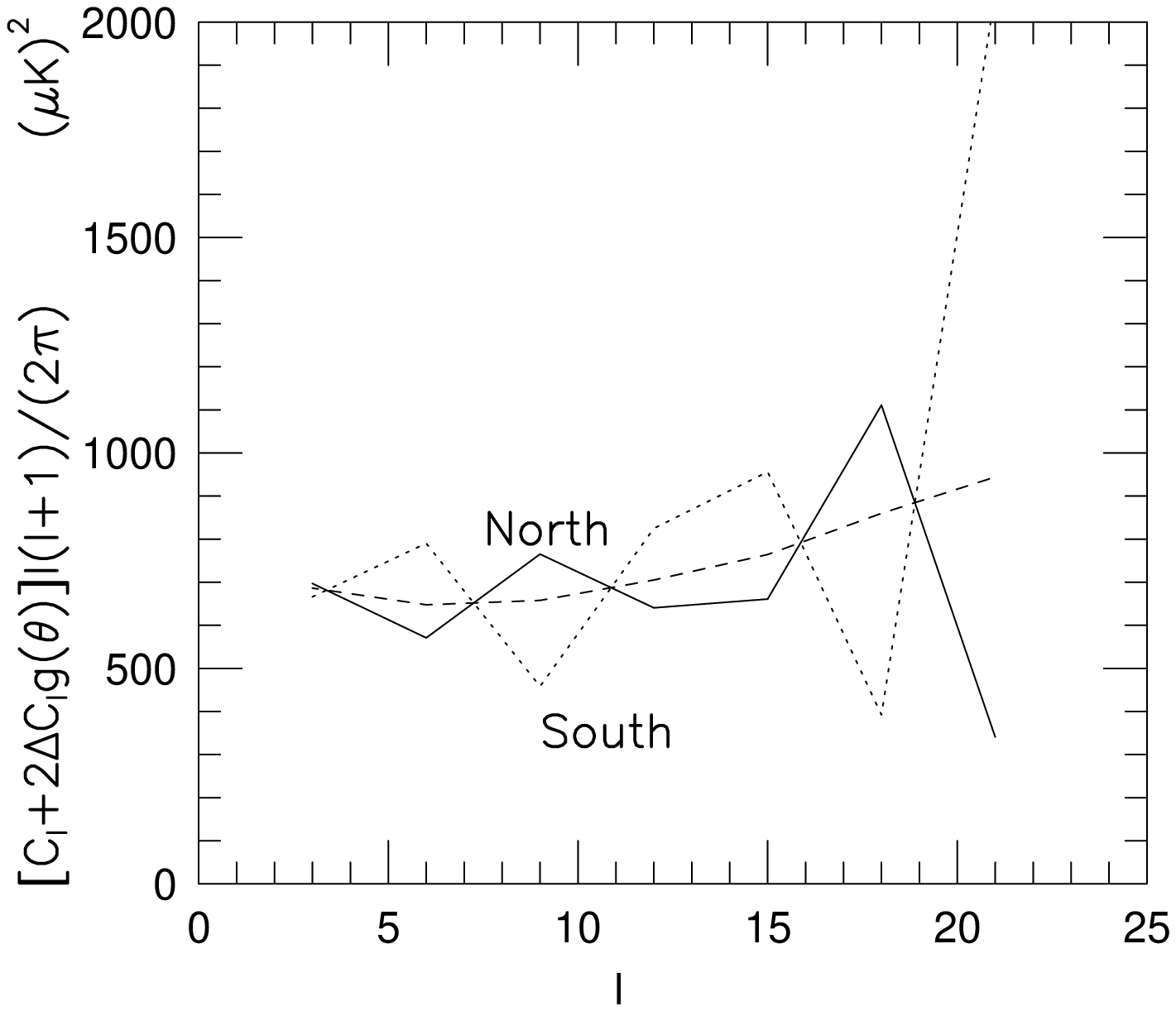}
\end{figure}

In Fig.~\ref{fig:erik}, the behaviors of $l(l+1) [C_l +2 \Delta C_l
g(\theta)] 
(\mu {\rm K})^2$ in the northern and southern poles are
shown for $\beta = -0.3$, in which we used the mean of four
types. We plotted $l(l+1) [C_l + 2 \overline{\Delta C_l} g(\theta)]$
for $l = 3, 6, 9, 
\cdots, 21$, where $\overline{\Delta C_l} \equiv (\Delta C_{l-1} +
\Delta C_l + \Delta C_{l+1})/3.$
For $R_0$, we adopted in the following $R_0 = -2.3 \times 10^{-5}$,
which was obtained as the best value in the previous paper. For $\beta
>$ and $< 0$, the deviation of \ $l(l+1) [C_l + 2 \Delta C_l
g(\theta)]$ in the 
northern pole from $l(l+1) C_l$ is larger and smaller than that in the
southern pole, respectively. It is found from this figure that our
model is at least qualitatively consistent with Fig.2 in
Eriksen et al.'s paper\cite{erk1}. In Table II, $\Theta_L/R_0, A/(R_0)^2,
B/(R_0)^2, C/(R_0)^2$ and $D/(R_0)^2$ are shown for $\beta = -0.3$. 

Next let us evaluate the scale-dependent estimators of skewness and
kurtosis. Since $\bar{S} (\theta)$ vanishes, we consider only $\bar{K} 
(\theta)$. As the essential property does not seem to depend on the
details of function $\psi (\theta,\theta_0)$, we assume here that it
is expressed as
\begin{equation}
  \label{eq:b2b}
\psi (\theta,\theta_0) = 1 \ {\rm for} \ 0 \leq \theta \leq
\theta_0, \ {\rm and} \ \psi (\theta,\theta_0) = 0 \ {\rm for} \ 
 \theta > \theta_0, 
\end{equation}
and that $N(\theta_0) = 1 - \cos \theta_0$.
Then  we obtain for $g(\theta) = \cos \theta + \beta \cos^2
\theta$ from Eq.(\ref{eq:m19}): 
\begin{equation}
  \label{eq:b3}
\bar{K} (\theta) =\zeta^2 \Bigl({\langle \Theta_P \Theta_{PL}
\rangle \over \langle \Theta_P \rangle^2}\Bigr)^2 (1-\mu)^2 \Bigl[1
+2\beta(1+\mu)+{4 \over 15}\beta^2 (4+7\mu+ 4\mu^2)\Bigr]/\Bigl\{1 +
\zeta \Bigl[1 +\mu +{2 \over 3}\beta (1 +\mu +\mu^2)\Bigr] \Bigr\}^2,   
\end{equation}
where 
\begin{equation}
  \label{eq:b4}
\zeta \equiv  {\langle \Theta_P \Theta_{PL} \rangle \over
\langle (\Theta_P)^2 \rangle}
\end{equation}
with $\mu = \cos \theta$. $\langle (\Theta_P)^2 \rangle$ and
$\langle \Theta_{PL} \rangle$ are derived from Eq.(\ref{eq:m12}).
In the case $\beta =0$, $\zeta$ is $-0.0624, -0.0143, -0.0666, -0.0222$
  for G, S, MG, MS, respectively. 
In the case $\beta =-0.3$, $\zeta$ is $-0.0787, -0.0352, -0.0813, -0.407$
  for G, S, MG, MS, respectively. 
The scale-dependent estimator of  kurtosis vanishes at $\theta = 0$
(the north pole) and 
increases with the increase of $\theta$. But it is of the order of
at most $4.0 \times 10^{-3} << 1$.

\begin{table}
\caption{CMB anisotropy powers $l(l+1) C_l$ and $l(l+1) \Delta
C_l$ in the case $\beta = -0.3$ and $n = 0.97$. The latter is caused by the 
coupling of cosmological perturbations and local inhomogeneities of
types G, S, MG and MS.\ Here $\xi \equiv 2\pi/[{\cal
P}_{F0} (T_0)^2]$, and ${\cal P}_{F0}$  and $R_0$ are the normalization 
factors.
\label{table1}}
\begin{tabular}{ccrrrrr}
\colrule
&\multicolumn{1}{c}{$l(l+1) C_l \xi$}&\multicolumn{4}{c}{$
10^{-3}\times 2l(l+1) \Delta C_l \xi/R_0 $}\\ 
$l$ & &G\ \ &S\ \ &MG\ \ &MS\ & \ mean \\
\colrule
$1$ & $4.550$ & $-0.78$ & $-1.98$ & $-0.76$ & $-1.43$ \ & $-1.24$\\ 
$2$ & $0.184$ & $-0.49$ & $-0.53$ & $-0.44$ & $-0.49$ \ & $-0.49$\\ 
$3$ & $0.177$ & $ 0.57$ & $-0.026$ & $ 0.75$ & $0.10$ \ & $ 0.35$\\ 
$4$ & $0.170$ & $-0.59$ & $-0.77$ & $-0.039$ & $-0.13$ \ & $-0.38$\\ 
$5$ & $0.168$ & $ 2.00$ & $ 0.18$ & $ 1.69$ & $ 0.22$ \ & $1.02$\\ 
$6$ & $0.166$ & $ 1.88$ & $ 0.82$ & $ 2.58$ & $ 1.86$ \ & $ 1.79$\\ 
$7$ & $0.167$ & $ 2.41$ & $ 1.20$ & $0.094$ & $-0.17$ \ & $0.88$\\ 
$8$ & $0.165$ & $ 3.09$& $ 1.37$ & $2.29$ & $ 0.60$ \ & $1.84$\\ 
$9$ & $0.172$ & $-5.14$& $-3.37$ & $-7.39$& $-3.73$ \ & $-4.91$\\ 
$10$ & $0.173$ & $-1.26$ & $-0.76$ & $-3.24$ & $-3.10$ \ & $-2.09$\\ 
$11$ & $0.179$ & $-5.01$ & $ 1.07$ & $-1.63$ & $3.50$ \ & $-0.52$\\ 
$12$ & $0.176$ & $-4.25$ & $-1.72$ & $-3.26$ & $0.51$ \ & $-2.18$\\ 
$13$ & $0.191$ & $5.76$& $ 3.63$& $10.61$& $3.20$ \ & $5.80$\\
$14$ & $0.191$ & $4.60$& $ 3.16$& $6.47$& $7.82$ \ & $5.51$\\
$15$ & $0.203$ & $7.21$ & $-9.53 $ & $4.47$ & $-9.62$ \ & $-1.87$\\ 
$16$ & $0.197$ & $ 2.39$ & $ 3.36 $ & $0.39$ & $-0.86$ \ & $1.32$\\ 
$17$ & $0.215$ & $-1.56$ & $10.28 $ & $-11.15$ & $3.57$ \ & $0.28$\\ 
$18$ & $0.220$ & $-4.54$ & $-1.61 $ & $-4.00$ & $-3.87$ \ & $-3.50$\\ 
$19$ & $0.230$ & $-19.91$ & $-8.12$ & $-8.58$ & $1.13$ \ & $-8.87$\\ 
$20$ & $0.231$ & $2.77$ & $-6.63$& $3.63$ & $ 2.08$ \ & $0.46$\\  
$21$ & $0.240$ & $33.138$ & $13.66 $ & $35.62$ & $13.42$ \ & $23.96$\\ 
$22$ & $0.260$ & $ 2.94$ & $13.74 $ & $-0.38$ & $1.95$ \ & $4.56$\\ 
\colrule
\end{tabular}
\end{table}

\begin{table}
\caption{CMB anisotropies caused by only the spot-like object of
types G, S, MG and MS. $R_0$ is the normalization factor.
\label{table2}}
\begin{tabular}{lccccc}
\colrule
model types &G&S&MG&MS& mean \\
\colrule
$\Theta_L/R_0$\ \ & $0.75$ & $0.022$ & $0.92$ & $0.23$& $0.48$\\
$A/(R_0)^2$\ \ & $2.5\times 10^4$& $1.5\times 10^4$ &
$3.0\times 10^4$ & $2.3\times 10^4$&$2.3 \times 10^4$ \\ 
$B/(R_0)^2$\ \ & $-9.1\times 10^2$& $-9.1\times 10^2$ & $-9.1\times
10^2$ & $-8.7\times 10^2$& $-9.0\times 10^2$\\ 
$C/(R_0)^2$\ \ & $31$& $45$ & $27$ & $36$& $35$\\ 
$D/(R_0)^2$\ \ & $10$& $15$ & $8.9$ & $11$& $11$\\ 
\colrule
\end{tabular}
\end{table}
%

\section{A cluster-scale spot-like object}
\label{sec:level4}

In this section we study the anisotropy of CMB radiation, coming
through the inside or neighborhood of an isolated cluster-scale
object, to consider non-Gaussian signatures of the spot-like
object \cite{viel,cruz}. The distance to the object is about $30h^{-1}$Mpc
and the angular radius is about 7 degree. 

The angular part of $F_L ({\bf x}) (= R(r) g(\theta))$ is assumed to be 
\begin{equation}
  \label{eq:d1}
g(\theta) = P_{l_1} (\cos \theta), \ {\rm for} \ 0 \le
\theta \le \theta_1, \ {\rm and} \
g(\theta) =  0  \ {\rm for} \ \theta > \theta_1, 
\end{equation}
respectively, with $l_1 = 20$, where $\cos \theta_1 = 0.992$ or
$\theta_1 = 
7.25 $deg.  The radial part $R(r)$ is assumed to have the four types
G, S, MG and MS (cf. Eq.(\ref{eq:b2}), in which $(x_1, x_2) = (r_1,
r_2)/r_c = (0.9, 1.1)$ and $(ar_c) \approx 30 h^{-1}$Mpc.

The powers of CMB anisotropies are calculated using
Eqs. (\ref{eq:m12}) and (\ref{eq:m13}) for $l = 1 \sim 40$, similarly
to those in Section III.  In Table III, the values of $l(l+1) 
\Delta C_l$ are shown for $l = 1 - 22$. It seems that there is an
oscillatory behavior with the period of $\Delta l \simeq 5$. In Table IV,
$\Theta_L/R_0, A/(R_0)^2, B/(R_0)^2, C/(R_0)^2$ and $D/(R_0)^2$ are shown.

Now let us derive the scale-dependent estimator of kurtosis 
\begin{equation}
  \label{eq:d2}
\bar{K} (\theta) =\zeta^2 \Phi (\theta)/[1 + 2\zeta \Psi (\theta)]^2,
\end{equation}
where $\zeta$ is defined by Eq.(\ref{eq:b4}) and

\begin{eqnarray}
  \label{eq:d3}
\Phi (\theta) &\equiv& 12 \Bigl\{{N(\theta,\theta_0)}^{-1} \int^1_0
 [g(\theta')]^2 \psi(\theta',\theta_0)d\mu' - \Psi^2 \Bigr\}, \cr
\Psi (\theta) &\equiv&  {N(\theta,\theta_0)}^{-1} \int^1_0
 g(\theta') \psi(\theta',\theta_0) d\mu'
\end{eqnarray}
with $\mu' = \cos \theta'$. The value of $\zeta/R_0$ is $-(4.8, 2.9,
2.4, 3.0) \times 10^6$ for G, S, MG and MS, respectively, and their
mean is $-3.2 \times 10^6$. As the fix point we adopt here the center
of the cluster-scale object.
    
To determine $R_0$, we consider here the first-order density
perturbation due to the cluster-scale object. It is expressed as
\begin{equation}
  \label{eq:d4}
(\mathop{\delta}_1 \rho/\rho)_L = {1 \over \rho a^2} \Bigl({a'\over a}P'
-1\Bigr) \tilde{R} P_{l_1} (\cos \theta)
\end{equation}
with $l_1 = 20$, where 
\begin{equation}
  \label{eq:d5}
\tilde{R} = (r_c)^{-2} [(x^2 R_{,x})_{,x}/x^2 - l_1(l_1+1) R/x^2].
\end{equation}
The ratio ($\delta M/M$) of perturbed mass ($\delta M$) to background
mass ($M$) in the interval $x = [x_1, x_2]$ is defined by
\begin{eqnarray}
  \label{eq:d6}
J &\equiv& \int^{x_2}_{x_1} \int^1_0 (\mathop{\delta}_1 \rho/\rho)_L x^2
dx d\mu/\Big\{{1 \over 3}[(x_2)^3 - (x_1)^3] \int ^1_0 d\mu \Big\} \cr
&=& {3 \over 2(\rho a^2)_0 (r_c)^2} \Bigl({a'\over a}P' -1\Bigr)_0
[(x_2)^3 - (x_1)^3]^{-1} \int^{x_2}_{x_1} [(x^2 R_{,x})_{,x} -
l_1(l_1+1) R] dx. 
\end{eqnarray}
The factor $[(a'/a)P' -1]_0$ is equal to $-0.456$ in the concordant
background model with $\Omega_0 = 0.27$ and $\Lambda_0 =
0.73$, and we have
\begin{equation}
  \label{eq:d7}
J = -3.9\times 10^3 ({30 h^{-1}{\rm Mpc} / a_0 r_c}
)^2 \int^{x_2}_{x_1} [(x^2 R_{,x})_{,x} -l_1(l_1+1) R] dx.
\end{equation}
After performing the integration in Eq.(\ref{eq:d7}), we obtain
\begin{equation}
  \label{eq:d8}
J = (2.5,0.94,3.1,1.9) \times 10^4 \  R_0\  ({30h^{-1}{\rm Mpc}/a_0
r_c})^2, 
\end{equation}
for types G, S, MG and MS, respectively, and their mean value is
$\bar{J} = 2.1 \times 10^4 \ R_0 \ ({30h^{-1}{\rm Mpc} /
a_0 r_c} )^2$. 

On the other hand, the observed value of $\langle (\delta M/M)^2
\rangle^{1/2}$ is about $1$ on the scale of $8h^{-1}$Mpc and for the
power spectrum $P(k) \propto k^n$, we have $\delta M/M \propto
M^{-(n+3)/6}  \propto r^{-(n+3)/2} = r^{-1.985}$ for $n = 0.97$.
Since we are considering a local inhomogeneity included in the sphere
of radius $2 a_0 r_c$, its scale is regarded as $4 \epsilon a_0 r_c$\
 ($\simeq 120 \epsilon h^{-1}$Mpc) with $\epsilon \approx 1$, so that
the value of $\delta M/M$ for the inhomogeneity is
\begin{equation}
  \label{eq:d9}
(\delta M/M)_{\rm power} = (8h^{-1}/4 \epsilon a_0 r_c)^{1.985} /b =
4.6\times 10^{-3} (b \epsilon^{1.985})^{-1} ({30h^{-1}{\rm Mpc} / a_0
r_c} )^{1.985}, 
\end{equation}
where $b$ is the biasing factor\cite{sut,lidd}. If we assume $|\bar{J}| =
(\delta M/M)_{\rm power}$, we obtain from Eq.(\ref{eq:d9}) 
\begin{equation}
  \label{eq:d11}
R_0 = 2.2 \times 10^{-7} (b \epsilon^{1.985})^{-1} ({30h^{-1}{\rm Mpc}
/ a_0 r_c})^{0.015}.  
\end{equation}
Accordingly it is reasonable to treat the case with $R_0 = (2 \sim
3)\times 10^{-7}$. 

\begin{figure}[t]
\caption{\label{fig:kurt} The angular dependence of the
scale-dependent estimator of kurtosis.} 
\includegraphics{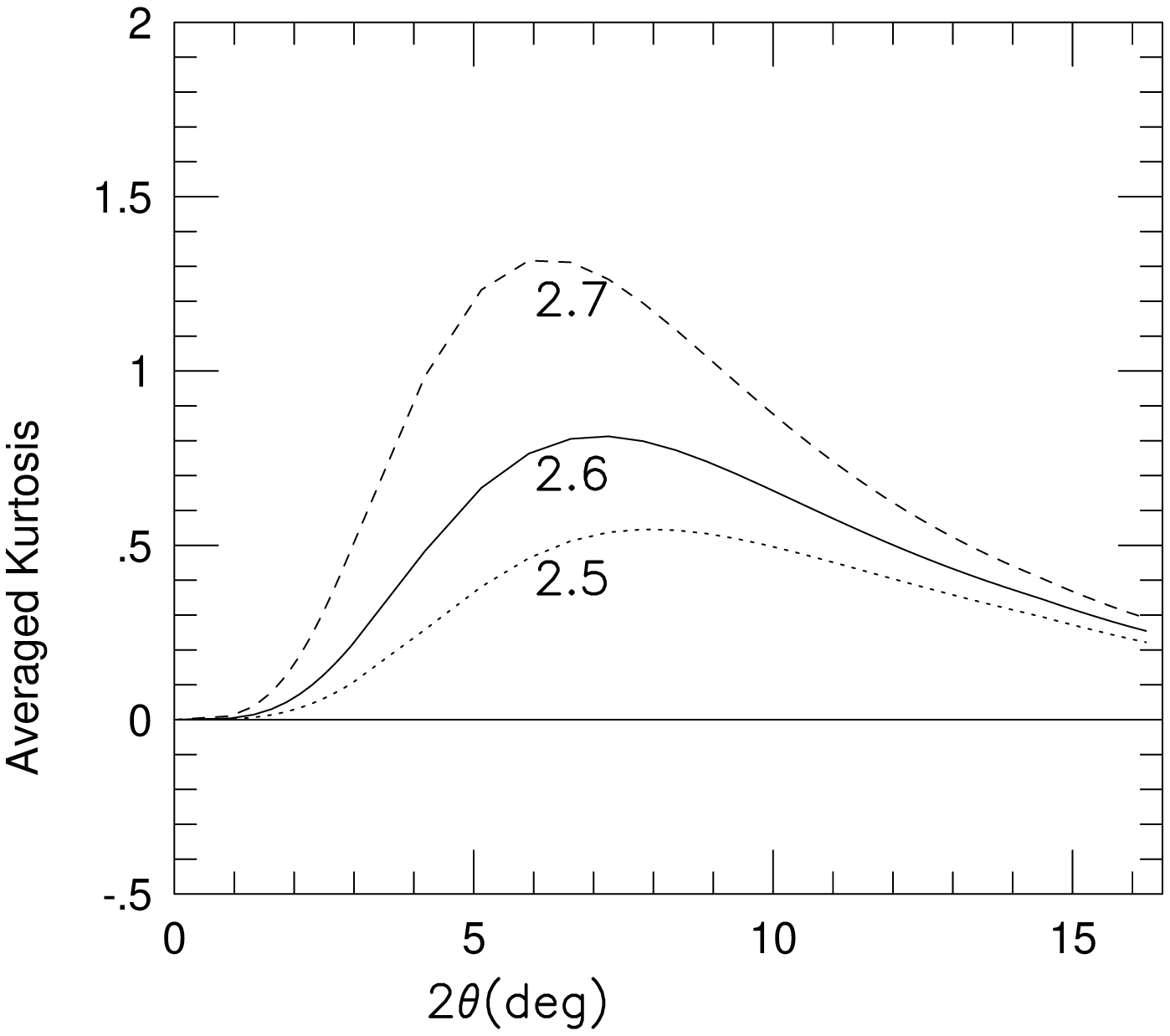}
\end{figure}

Now let us show the behavior of the scale-dependent estimator of
skewness and kurtosis. Since the skewness estimator vanishes, we
consider only $\bar{K} (\theta)$ for $R_0 = (2.5, 2.6, 2.7)\times
10^{-7}$. Here we assume the simple functional form Eq.(\ref{eq:b2b})
for $\psi (\theta,\theta_0)$ and $N(\theta_0) = 1 - \cos \theta_0$ in
a similar way to the previous section. Then the behavior of $\bar{K}$
is shown 
in Fig.~\ref{fig:kurt} and the peak of the kurtosis is found to appear
around $\theta \approx 3.5$ deg, and so the diameter 
$(2 \theta)$ of the circular region with the peak is about $7$
deg. The estimator of dispersion $\bar{\sigma} (\theta)$ (being the
denominator of Eq.(\ref{eq:d2})) decreases and so $\bar{K}
(\theta)$ increases with the increase of $R_0$ around these values.
The values of the kurtosis at the peaks are $1.3, 0.81, 0.55$ for 
$R_0 = (2.5, 2.6, 2.7) \times 10^{-7}$, respectively. It is therefore
found from this figure that the case $R_0 = 2.6 \times 10^{-7}$ is 
best to represent the observed kurtosis \cite{viel,cruz}.  

Thus we found that the scale-dependent estimator of kurtosis does not
vanish, though kurtosis in the original definition vanishes. The
origin of this strange situation may be in the definition of the
estimator (Eq.(\ref{eq:m18})) and Vielva et al's definition (Eq.(18) in
\cite{viel}) which are given as the ratio of the
two expectation values.   


%
\begin{table}
\caption{CMB anisotropy powers $l(l+1) C_l$ and $l(l+1) \Delta
C_l$ in the cluster-scale case $n = 0.97$. The latter is caused by the 
coupling of cosmological perturbations and the cluster-scale inhomogeneity of
types G, S, MG and MS.\ Here $\xi \equiv 2\pi/[{\cal
P}_{F0} (T_0)^2]$, and ${\cal P}_{F0}$  and $R_0$ are the normalization 
factors.
\label{table3}}
\begin{tabular}{ccrrrrr}
\colrule
&\multicolumn{1}{c}{$l(l+1) C_l \xi$}&\multicolumn{4}{c}{$
10^{-5}\times 2l(l+1) \Delta C_l \xi/R_0 $}\\ 
$l$ & &G\ \ &S\ \ &MG\ \ &MS\ & \ mean \\
\colrule
$1$ & $4.550$ & $-304.8$ & $-173.8$ & $-135.1$ & $-171.3$ \ & $-196.3$\\ 
$2$ & $0.184$ & $-7.96$ & $-4.71$ & $-3.81$ & $-4.84$ \ & $-5.33$\\ 
$3$ & $0.177$ & $-10.44$ & $-6.32$ & $-5.15$ & $-6.51$ \ & $-7.11$\\ 
$4$ & $0.170$ & $-11.94$ & $-7.43$ & $-6.17$ & $-7.77$ \ & $-8.33$\\ 
$5$ & $0.168$ & $-19.05$ & $-11.40$ & $-9.04$ & $-11.34$ \ & $-12.71$\\ 
$6$ & $0.166$ & $-17.96$ & $-11.24$ & $-9.11$ & $-11.33$ \ & $-12.41$\\ 
$7$ & $0.167$ & $-1.60$ & $-1.06$ & $-0.70$ & $-0.73$ \ & $-1.02$\\ 
$8$ & $0.165$ & $-6.36$& $-4.27$ & $-3.30$ & $-3.90$ \ & $-4.46$\\ 
$9$ & $0.172$ & $-4.19$& $-3.14$ & $-2.84$& $-3.50$ \ & $-3.42$\\ 
$10$ & $0.173$ & $-1.62$ & $-1.41$ & $-1.32$ & $-1.54$ \ & $-1.47$\\ 
$11$ & $0.179$ & $-17.75$ & $-10.95$ & $-8.80$ & $-10.93$ \ & $-12.11$\\ 
$12$ & $0.176$ & $-16.40$ & $-9.29$ & $-7.29$ & $-9.22$ \ & $-10.55$\\ 
$13$ & $0.191$ & $5.74$& $ 2.75$& $2.01$& $2.70$ \ & $3.30$\\
$14$ & $0.191$ & $13.41$& $ 7.34$& $5.20$& $6.58$ \ & $8.13$\\
$15$ & $0.203$ & $-8.28$ & $-6.40 $ & $-5.98$ & $-7.38$ \ & $-7.01$\\ 
$16$ & $0.197$ & $-22.76$ & $-14.10 $ & $-11.73$ & $-14.67$ \ & $-15.82$\\ 
$17$ & $0.215$ & $-17.28$ & $-8.55 $ & $-5.74$ & $-7.30$ \ & $-9.72$\\ 
$18$ & $0.220$ & $13.28$ & $9.04 $ & $7.68$ & $9.71$ \ & $9.93$\\ 
$19$ & $0.230$ & $27.03$ & $16.66$ & $12.82$ & $15.77$ \ & $17.32$\\ 
$20$ & $0.231$ & $-12.86$ & $-5.15$& $-3.62$ & $-5.02$ \ & $-6.67$\\  
$21$ & $0.240$ & $-35.12$ & $-22.95 $ & $-19.08$ & $23.70$ \ & $-13.36$\\ 
$22$ & $0.260$ & $-0.663$ & $-1.99 $ & $0.350$ & $0.614$ \ & $-0.42$\\ 
\colrule
\end{tabular}
\end{table}

\begin{table}
\caption{CMB anisotropies caused by only the spot-like object of
types G, S, MG and MS. $R_0$ is the normalization factor.
\label{table4}}
\begin{tabular}{lccccc}
\colrule
model types &G&S&MG&MS& mean \\
\colrule
$\Theta_L/R_0$\ \ & $179.$ & $-0.104$ & $1.66$ & $0.313$& $45.3$\\
$A/(R_0)^2$\ \ & $-9.1\times 10^8$& $-4.2\times 10^8$ &
$-3.1\times 10^8$ & $-4.4\times 10^8$&$-4.2 \times 10^8$ \\ 
$B/(R_0)^2$\ \ & $-1.4\times 10^5$& $-1.5\times 10^5$ & $-1.8\times
10^5$ & $-1.7\times 10^5$& $-1.6\times 10^5$\\ 
$C/(R_0)^2$\ \ & $3.1\times 10^{-3}$& $3.5\times 10^{-4}$ & $2.8\times
10^{-4}$ & $3.9\times 10^{-4}$& $1.0\times 10^{-3}$\\ 
$D/(R_0)^2$\ \ & $6.3\times 10^3$& $2.6\times 10^3$ & $2.0\times
10^3$ & $2.7\times 10^3$& $3.4\times 10^3$\\ 
\colrule
\end{tabular}
\end{table}
%

\section{Concluding remarks}
\label{sec:level5}

The large-scale local inhomogeneity with the north-south asymmetry
was found to reproduce the observed asymmetry of CMB anisotropies in
the $l(l+1)(C_l + 2\Delta C_l g(\theta))$ diagram in
the case when not only a dipole component but also a small quadrupole
component are included. It is interesting that through the second-order
nonlinearity, the asymmetry in the
CMB anisotropies observed in WMAP may be related to the asymmetry in a
large-scale structure observed in SDSS. 

In our models of large-scale and cluster-scale inhomogeneities, the
skewness and kurtosis in the 
original definition vanish, but a non-Gaussian signal similar to
that observed in the spot-like object\cite{viel} was found to be
theoretically derived in our model by using the scale-dependent
estimator of kurtosis. It may, therefore, be difficult to conclude the
existence of pure non-Gaussianity from obtaining observationally the
non-vanishing values of scale-dependent estimators of skewness and
kurtosis.  

In the large-scale and cluster-scale cases,
the dependence of our results on $a r_c$ is very small, and that on
$x_1$ and $x_2$ also is qualitatively small, though quantitatively
sensitive.  

It is emphasized finally that the local inhomogeneities we considered
in our models are not any extraordinary outliers, but usual
inhomogeneities with ordinary mass spectra, and that their important
unique character may be so near and isolated that they can be regarded
as being not included in 
the primordial random perturbations. In the cluster-like object, it is
additionally required that the dispersion ($\bar{\sigma}$) is
comparatively small, to show a remarkable spot-like appearance.  

\begin{acknowledgments}
The author thanks E. Komatsu for helpful discussions on the harmonic
expansion of second-order anisotropies. He also thanks K.T. Inoue for
various discussions on the observed asymmetry of
CMB anisotropies. Numerical computation in this work was carried out
at the Yukawa Institute Computer Facility.  
\end{acknowledgments}

\appendix
\section{Derivation of $\Theta_{LL}$}
\begin{equation}
  \label{eq:a1}
\Theta_{LL} = A g^2 + B (g')^2 + C g(g'' +\cot \theta g') + D
[(g'')^2 + \cot^2 \theta (g')^2], 
\end{equation}
where
\begin{eqnarray}
  \label{eq:a2}
A &=& {1 \over 8}\Bigl(\int^{\lambda_e}_{\lambda_o}
d{\lambda} P' R_{,rr} \Bigr)^2 - {1 \over 4}
\int^{\lambda_e}_{\lambda_o} d{\lambda} PP'(R_{,rr})^2 \cr
&+& {1 \over 4} \int^{\lambda_e}_{\lambda_o}
d{\lambda} P''R_{,rr}  \int^{\lambda}_{\lambda_o} d{\bar{\lambda}}
P(\bar{\eta}) R_{,rr} (\bar{r}) \cr
&+& {1 \over 4} \int^{\lambda_e}_{\lambda_o}
d{\lambda} \Bigl\{-P'\Bigl[RR_{,rr} +{5 \over 4}(R_{,r})^2\Bigr] \cr
&+& {1 \over 14} PP'[19(R_{,rr})^2  -12 \tilde{R} R_{,rr} -3(R_{,rr}
+\tilde{R})(R_{,rr} -\tilde{R}) -6(R_{,r}/r)^2]\}, \cr
B &=& {1 \over 4}\int^{\lambda_e}_{\lambda_o} d{\lambda} \{{1 \over 4}
P'R^2/r^4 
+ {1 \over 14} PP' [19(R_{,r}/r)^2 - 6(R_{,r}/r - R/r^2)]\}, \cr
C &=& -{3 \over 28} \int^{\lambda_e}_{\lambda_o} d{\lambda} PP'
RR_{,r}/r^3, \cr
D &=& -{3 \over 56} \int^{\lambda_e}_{\lambda_o} d{\lambda} PP' (R/r^2)^2.
\end{eqnarray} 
%


\end{document}